\documentclass[aps, pre, twocolumn,showkeys,showpacs,amsmath,amssymb,superscriptaddress]{revtex4-1}

\usepackage[T1]{fontenc}
\usepackage[latin1]{inputenc}

\usepackage[ngerman,british]{babel}
	\addto\captionsngerman{}

\usepackage{amsmath,amssymb,amsthm,dsfont,mathtools,mathrsfs,bbm}

\usepackage{mdwlist}
\usepackage{paralist}
\usepackage{array}
\usepackage{booktabs}						
\usepackage{dcolumn} 						

\usepackage{graphicx}
\usepackage{wrapfig}

\usepackage{url}

\usepackage{color}
	\definecolor{myblue}{rgb}{0,0.3,0.8}
	\definecolor{mygreen}{rgb}{0,0.5,0}
	\definecolor{myblue}{rgb}{0,0.3,0.8}

\usepackage[version=3]{mhchem}

\usepackage{savesym}

\savesymbol{a}
\savesymbol{b}
\savesymbol{d}
\savesymbol{e}
\savesymbol{k}
\savesymbol{l}
\savesymbol{H}
\savesymbol{L}
\savesymbol{p}
\savesymbol{r}
\savesymbol{s}
\savesymbol{S}
\savesymbol{F}
\savesymbol{t}
\savesymbol{w}
\savesymbol{th}
\savesymbol{div}
\savesymbol{tr}
\savesymbol{Re}
\savesymbol{Im}
\savesymbol{AA}
\savesymbol{SS}
\savesymbol{Si}
\savesymbol{gg}
\savesymbol{ss}

\newcommand{\a}{\alpha}
\newcommand{\b}{\beta}
\newcommand{\e}{\varepsilon}
\newcommand{\d}{\delta}
\newcommand{\del}{\partial}

\newcommand{\g}{\gamma}
\newcommand{\G}{\Gamma}

\newcommand{\r}{\rho}
\newcommand{\s}{\sigma}

\newcommand{\w}{\omega}

\newcommand{\cov}{{\nabla}}

\newcommand{\R}{\mathbbm{R}}
\newcommand{\C}{\mathbbm{C}}

\newlength{\tmplen}

\usepackage{bbold}

\begin{document}

\title{Landau levels in wrinkled and rippled graphene sheets}

\author{Kyriakos Flouris}\affiliation{ %
ETH
  Z\"urich, Computational Physics for Engineering Materials, Institute
 for Building Materials, Wolfgang-Pauli-Str. 27, HIT, CH-8093 Z\"urich,
 Switzerland}%
\author{Miller Mendoza Jimenez}\affiliation{ %
ETH
  Z\"urich, Computational Physics for Engineering Materials, Institute
 for Building Materials, Wolfgang-Pauli-Str. 27, HIT, CH-8093 Z\"urich,
 Switzerland}%
\author{Hans J. Herrmann}
\affiliation{ %
ETH
  Z\"urich, Computational Physics for Engineering Materials, Institute
 for Building Materials, Wolfgang-Pauli-Str. 27, HIT, CH-8093 Z\"urich,
 Switzerland}%
 \affiliation{ %
Departamento de F\' isica, Universidade do Cear\' a, 60451-970 Fortaleza, Brazil}%
  \affiliation{ %
on leave from C.N.R.S., UMR 7636, PMMH, ESPCI, 10 rue Vauquelin, 75231 Paris Cedex 05, France }%

\begin{abstract}%
We study the discrete energy spectrum of curved graphene sheets in the presence of a magnetic field. The shifting of the Landau levels is determined for complex and realistic geometries of curved graphene sheets. The energy levels follow a similar square root dependence on the energy quantum number as for  rippled and flat graphene sheets. The Landau levels are shifted towards lower energies proportionally to the average deformation and the effect is larger  compared to a simple uni-axially rippled geometry. Furthermore, the resistivity of wrinkled graphene sheets is calculated for different average space curvatures and shown to obey a linear relation. The study is carried out with a quantum lattice Boltzmann method, solving the Dirac equation on curved manifolds.


\end{abstract}

\maketitle

\section{Introduction}

The re-discovery of graphene in 2004 has initiated a plethora of research around this 'wonder' material.\cite{illario,oliver1,oliver2,miller1,miller2} The characteristic 2D honeycomb carbon atom lattice makes it a perfect flat electronic material which can be stacked and reshaped resulting in spectacular electronic properties.\cite{graphenerev1,cdp} As a first approximation graphene charge carriers act like mass-less quantum relativistic particles with the characteristic touching Dirac cones band structure $E=v_F \hbar k$, where E is the energy, $v_f$ the Fermi velocity, and $\hbar$ is the Planck constant. The electronic band structure of graphene is well-described by the tight-binding Hamiltonian and by approximating the electronic system via a superposition of local wave functions for isolated atoms.\cite{graphene_tb1} 
In reality graphene sheets are not perfectly flat, but instead have out of plane deformations mainly in the form of ripples and wrinkles.\cite{graphene_ripples} Extending to curved space can provide a more realistic description of this material.\cite{OLIVALEYVA} 

It has been recently shown that the quantified electronic energy levels in a magnetic field are shifted by curvature effects in strained graphene sheets.\cite{jd_paper} This work was restricted to a simple, effectively one-dimensional rippled manifold. In this work we build on this result and simulate a more complex, diverse and realistic graphene geometry. More specifically this is to include wrinkles and two dimensional ripples as seen experimentally Ref.~\cite{graphene_wrinkles_ripples}.  

As analytic solutions of the Dirac equation in complex curved space are difficult if not impossible to find in this work numerical solutions based on the quantum lattice Boltzmann (QLB) method\cite{succi_qlbm} are obtained. The method is extended to curved spaces\cite{miller_rlbm,fsi_flouris} and related to strained graphene by exploiting the similarities between the curved Dirac Hamiltonian and the effective Hamiltonian for strained graphene as done in Ref.~\cite{jd_paper}. This opens up the possibility of exporting the proven advantages of the curved space lattice Boltzmann methods, such  as computational efficiency, parallelizability and easy handling of complex geometries, to graphene.

In the next section the theories of curved space Dirac equation and strained graphene are explained. Followed by the method and results, where the Landau level shifting and the resistivity for the different geometries are presented. The results are summarized in the final section.

\section{Theory of strained graphene}

 The original Dirac Equation Ref.~\cite{dirac} can be naturally extended to curved space described by a metric tensor $g_{\mu\nu}$ by minimally coupling with a covariant derivative as 
\begin{equation}
\label{eq:Dirac}
(i \g^{\mu}D_{\mu}-m)\Psi=0,
\end{equation}
 in natural units such that $\hbar=c=1$ for $\hbar$, Planck's constant and $c$, the speed of light, where $\mu={0,1,2}$ for 2D space-time. $\Psi = (\Psi_a^+, \Psi_a^-) = (\psi_1^+,\psi_2^-,\psi_1^-,\psi_2^+) \in \C^4$ denotes the Dirac spinor for particle, hole $+,-$, and $\g^\mu = \g^\a e_\a^{~\mu}$ are the generalized space dependent  $\g$-matrices, where $\g^\a \in \C^{4\times 4}$ are the standard Dirac matrices, $e_\a^{~\mu} $ is the tetrad, which relates the flat Minkowski space ($\alpha$) to the curved space-time ($\mu$).  Here the tetrad is defined by $e_{\alpha}^{~\mu} g_{\mu \nu}e_{\beta}^{~\nu}=\eta_{\alpha \beta}$, where $g_{\mu \nu}$ denotes the metric tensor and $\eta_{\alpha \beta}$ is the 
Minkowski metric. 
  $D_{\mu}$, the covariant spinor derivative defined by $D_{\mu} \Psi=\del_{\mu} \Psi + \G_{\mu} \Psi$, where $\G_{\mu}$ denotes the spin connection matrices given by $\Gamma_\mu = - i/4 \w_\mu^{\a\b} \s_{\a\b}$,
with $\s_{\a\b} = i/2 [\g_\a,\g_\b]$ and $\w_\mu^{\a\b}= e_\nu^{~\a} \cov_\mu e^{\nu \b}$.



To model the single layer carbon atom honeycomb lattice structure we start from the tight binding Hamiltonian which is constructed assuming superposition of local wave-functions for isolated atoms on a honeycomb lattice.\cite{graphene_tb1} In the low energy limit it has been shown that the tight binding Hamiltonian converges to the Dirac Hamiltonian in the continuum limit,
\begin{equation}
\label{eq:dirac_hamiltonian}
H_D=-i v_f \int \Psi^\dagger \gamma^0 \gamma^i \del_i \Psi d^2x, 
\end{equation}
in natural units, where $\Psi$ is in the chiral representation, $v_f$ is the Fermi velocity.  In the context of graphene, the general Dirac spinor is defined as $\Psi=(\Psi_a^K,\Psi_a^{K'})=(\psi_A^K,\psi_B^{K'},\psi_A^{K'},\psi_B^{K})$, for sub-lattices $A,B$ and valleys $K,K'$.
The convergence from the tight binding Hamiltonian to  Eq.~(\ref{eq:dirac_hamiltonian}) can be seen as the Dirac cones in graphene with linear dispersion relation at the conduction and valence band connecting point $E=p$ for $E$, energy and $p$, momentum.

The equation of motion from this Hamiltonian is simply the  Dirac equation. 
In this work, we consider a static space-time metric with trivial time components
\begin{align*}
	g_{\mu\nu} = 
	\begin{pmatrix}
		1 & 0 \\
		0 & -g_{ij}	
	\end{pmatrix},
\end{align*}
where the latin indices run over the spatial dimensions. This simplifies the Dirac equation Eq.~(\ref{eq:Dirac})
to
\begin{equation}
\del_t \Psi + \s^a e_a^{~i}(\del_i + \G_i)\Psi = 0-i\g^0 m \Psi,
\end{equation}
with $\s^a = \g^0 \g^a$. After addition of external vector and scalar potentials $A_i(x)$ and $V(x)$ respectively as explained in Ref.~\cite{jd_paper}, the Dirac equation takes the following form:
\begin{equation}
\label{eq:Diracfull}
\del_t \Psi + \s^a e_a^{~i}(\del_i + \G_i- i A_i)\Psi =-i\g^0 (m-V) \Psi.
\end{equation}
Defining the Dirac current with $J^\mu = \overline \Psi \g^\mu \Psi$, the conservation law for can be written as $\del_t \rho + \nabla_i J^i = 0$,
where $\r = \Psi^\dagger \Psi \in \R$ and the $J^i = \overline{\Psi}\gamma^i \Psi \in \R$.

The standard Dirac Hamiltonian for Eq.~(\ref{eq:Diracfull} equation is
\begin{equation}
\label{eq:hamiltoniandirac}
H_D=-i \int \Psi^\dagger \sigma^a e_a^{~i}( \del_i + \G_i-i A_i)\Psi \sqrt{g} d^2x,
\end{equation}
For graphene the effective Hamiltonian looks like:\cite{OLIVALEYVA}
\begin{equation}
\label{eq:graphene_hamiltonian}
H^*_D=-i v_f \int \Psi^\dagger \sigma^a (v_a^{*i} \del_i + \G_a^*-i A^*_a)\Psi d^2x,
\end{equation}
where $v_a^{* i}=\d_{a i} + u_{a i} -\beta \e_{a i}$ is the space depended Fermi velocity, $\G^*_a=\frac{1}{2 v_f}\del_j v_a^{* j}$ is a complex gauge vector field which guarantees the hermiticity of the Hamiltonian and $A^*_a$ is a strain-induced pseudo-vector potential given by $A^*_a=(A_x^*,A_y^*)=\frac{\beta}{2a}(\e_{xx}-\e_{yy},-2\e_{xy}$), $\beta$ is the material dependent electron Grueneisen parameter, $a$ the lattice spacing and $\e_{i\jmath}= u_{i\jmath} +\frac{1}{2}\del_i h \del_j h$ the general strain tensor with in-plane, $u_{i\jmath}$ and out of plane, $h$ deformations.The term $u_{a i}$ in $v_a^{* i}$ can be interpreted as the deformation potential term and is purely a geometric consequence due to lattice distortion, it does not depend on the material as long as it has the same topology. Comparing this to the standard Dirac Hamiltonian in curved space Eq.~(\ref{eq:hamiltoniandirac})
we can match both Hamiltonians $H_D$ and $H^*_D$ by fulfilling the following relations:
\begin{equation}
\label{eq:effectivefields}
v_a^{*i}= \frac{1}{v_f} \sqrt{g}e_a^{~i}, \ \ \G_a^*= \frac{1}{v_f} \sqrt{g}e_a^{~i}\G_i, \ \ A^*_a= \frac{1}{v_f} \sqrt{g}e_a^{~i}A_i.
\end{equation}
All three can be simultaneously fulfilled by an effective metric tensor derived from the explicit expression of the tetrad.\cite{jd_paper} The effective Dirac model for non-uniformly strained graphene, as explained in Ref.~\cite{OLIVALEYVA}, relies on the basic principle that \textit{the theory for graphene under nonuniform strain should describe the particular case of a uniform strain},\cite{olivaleyva2} where both Dirac points in the Brillouin zone are affected equivalently such that the shifts of the valleys K and K' are equal in magnitude but have opposite signs. When moving from a flat to a rippled graphene sheet there is a transition from two to three-dimensional space. The strained graphene theory accounts for the change by effectively adjusting the hopping parameter relative to the particle proximity. The effect is then mapped to three-dimensional space through the relations in Eq.~(\ref{eq:effectivefields}).
The numerical solutions are obtained with the Quantum Lattice Boltzmann Method as described in Ref.~\cite{jd_paper}. 

\section{Method and results}
\begin{figure*}
\begin{center}
\includegraphics[width=2\columnwidth]{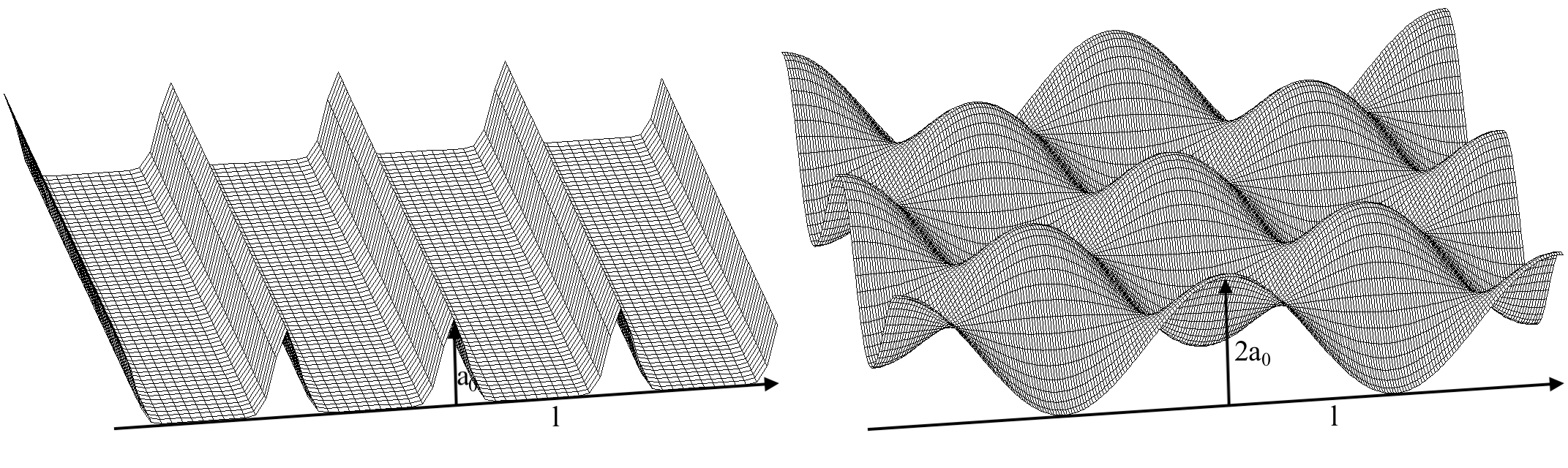}
\caption{\label{fig:geometries} Left: wrinkled graphene sheet geometry. Right: Bi-axially rippled graphene sheet geometry}
\end{center}

\end{figure*}

\subsection{Shifted Landau levels in curved graphene sheets \label{sec:LL}}

To investigate curved graphene sheets the computational domain is initialized as a two-dimensional out of plane rippled surface as shown in the left of Fig.~\ref{fig:geometries}. Defining  $\mathbf{r}=(x,y,h(x,y))$, where $(x,y)$ are the plane coordinates an h represents the out of plane deformations with 
\begin{equation}
\label{eq:wrinkles}
    h(x,y)=a_0 \cos\bigg(\frac{2\pi k_0x}{l}\bigg)^\eta,
\end{equation}
where $a_0$ is the amplitude $k_0$ a measure of the frequency and $\eta \in 2\mathbb{Z^+}$ controls the sharpness of the wrinkles, in this work we use $\eta=20$, see Fig.~\ref{fig:geometries}. This implies a metric tensor 
\begin{align*}
	g_{\mu\nu} = 
	\begin{pmatrix}
		1+h_x^2 & 0 \\ 
		0 & 1	
	\end{pmatrix},
\end{align*}
where $h_x:=\del_x h(x,y)$. A magnetic field $B$ is applied along the the out of plane perpendicular direction $z$. In this Landau gauge the vector potential in Cartesian coordinates is given by $A=(A_x,A_y, A_z)=(0,Bx)$. This corresponds to a magnetic field $\nabla \times A = (0~0~ B)^T$. Furthermore, defining $f(x)=0.5 h_x^2$, the position dependent Fermi velocity,  $v^*$  the generalized strain tensor,  $\epsilon^*$,  the complex vector field $\Gamma^*$ and the pseudo-vector potential $A^*$ are given by
\begin{align*}
	\epsilon_{ij} = 
	\begin{pmatrix}
	    f(x) & 0 \\ 
		0 & 0	
	\end{pmatrix},
	~	
	v^{*i}_a = 
	\begin{pmatrix}
		1-f(x) & 0 \\ 
		0 & 1	
	\end{pmatrix},      
	\\
    \Gamma^*=
     \begin{pmatrix}
	-\frac{f'(x)}{2} & 0 \\ 
	\end{pmatrix}, 	
     ~ 
     A^*_a=(0,Bx)
\end{align*}
implying according to Eq.~(\ref{eq:effectivefields}) that the corresponding effective metric tensor is
\begin{align*}
	g_{ij} = 
	\begin{pmatrix}
     	1 & 0 \\ 
		0 & (1-f(x))^2	
	\end{pmatrix}.
\end{align*}
The wave-function is initialized to a zero momentum gaussian wave-packet by 
  \begin{align}
  \label{eq:wavefunction}
    \Psi(t=0)= \frac{1}{2\sqrt{4 \pi}}
         & \begin{pmatrix}
           1 \\
           0 \\
           0 \\
           i
         \end{pmatrix}
         e^{-\frac{x^2}{4}}
         = 
          \frac{1}{2\sqrt{4 \pi}}
           \begin{pmatrix}
           1 \\
           \lambda i
         \end{pmatrix}
         e^{-\frac{x^2}{4}},
\end{align}
in the Dirac spinor basis and the graphene basis respectively,\cite{cdp} where $\lambda=\pm 1$ is the band index.

To measure the energy levels of the system the wavefunction can be decomposed into an infinite sum of the energy eigenfuctions $\Psi_n$ with eigenvalues $E_n$. From the time evolution of the eigen-functions, $\Psi_n(t)=\Psi_n(0)\exp(-iE_nt)$, the time evolution of the complete Dirac field is given by:
\begin{equation}
    \Psi(t)=\sum_{n\in\mathbb{Z}}a_n\Psi_n(t)=\sum_{n\in\mathbb{Z}}a_n\Psi_n(0)e^{iE_nt},
\end{equation}
where $a_n$ corresponds to the coefficient of the $n^{th}$ energy eigen-function, which can be determined by its overlap relative to $\Psi(t=0)$. The energy levels will correspond to the Landau levels ($E_n$) and can be determined by a standard Fourier transform of the time evolution of the space dependent spinor:
\begin{equation}
\label{eq:FFT}
\begin{split}
    \mathcal{F}[\Psi](E)  & =\int \Psi(t)e^{iEt}dt  \\
  & =\sum_{n\in\mathbb{Z}}a_n\Psi_n(0) \cdot \int e^{i(E-E_n)t} dt \\
  & =\sum_{n\in\mathbb{Z}}a_n\Psi_n(0) \cdot 2 \pi \delta(E-E_n). 
\end{split}
\end{equation}
For special relativistic fermions and  flat graphene sheets the quantized Landau levels follow the established result:
\begin{equation}
\label{eq:llflat}
    E_n=sgn(n)\sqrt{2B|n|},
\end{equation}
where $n$ corresponds to the energy quantum number.

\begin{figure*}
\begin{center}
\includegraphics[width=2\columnwidth]{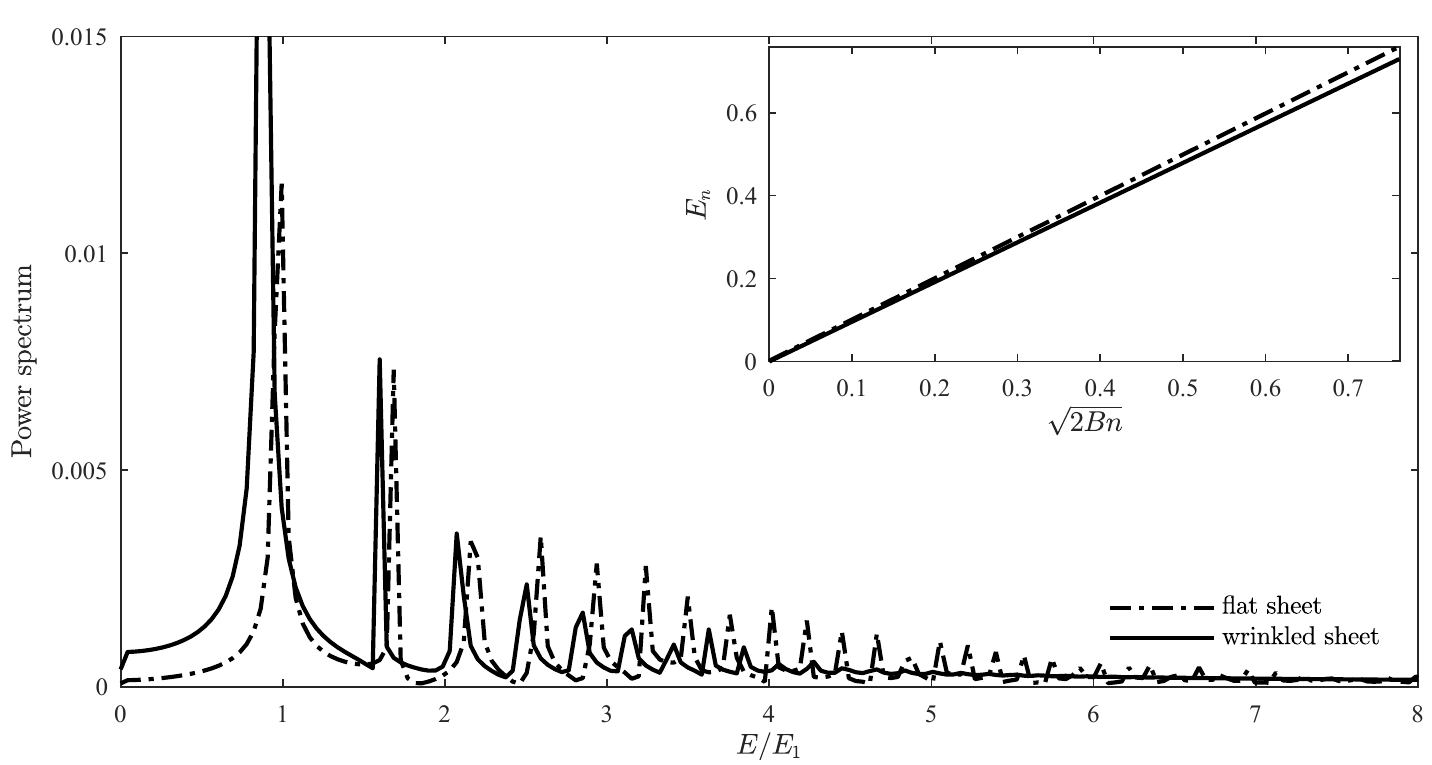}
\caption{\label{fig:FFT} Power spectrum of discrete energy levels in the presence of a magnetic field, B=0.01 or 1T in real units, for wrinkled and flat graphene sheet. The energy axis is normalized to the first energy level $E_1$ which is $\sim35$meV. Inset shows the dependence of the  peak position on the square root of the magnetic field for wrinkled and flat graphene sheets. For flat graphene sheets the slope is  equal to one. }
\end{center}
\end{figure*}

This relation is affected by curvature.\cite{jd_paper} For the wrinkled geometry we consider a periodic lattice of length $l_x=0.2 \mu m$, discretized into $L_x \times L_y=512 \times 128$ grid points and apply a magnetic field $B=0.01$. The wave function is initialized as in Eq.~(\ref{eq:wavefunction}). Figure \ref{fig:FFT} depicts the energy spectrum for the flat case relative to  the wrinkled geometry as described by Eq.~(\ref{eq:wrinkles}). The energy spectrum is obtained from the time dependence of the wave-function via the Fourier transform, Eq.~(\ref{eq:FFT}). The Landau levels are seen as the discrete peaks corresponding to $n=1,3,5...$. Only the odd energy states are excited due to the wave-function being symmetric and no anti-symmetric eigenstates are initialized. The wrinkled graphene Landau levels are different in width and amplitude relative to the flat graphene implying different eigenstates.

\begin{figure*}
\begin{center}
\includegraphics[width=2\columnwidth]{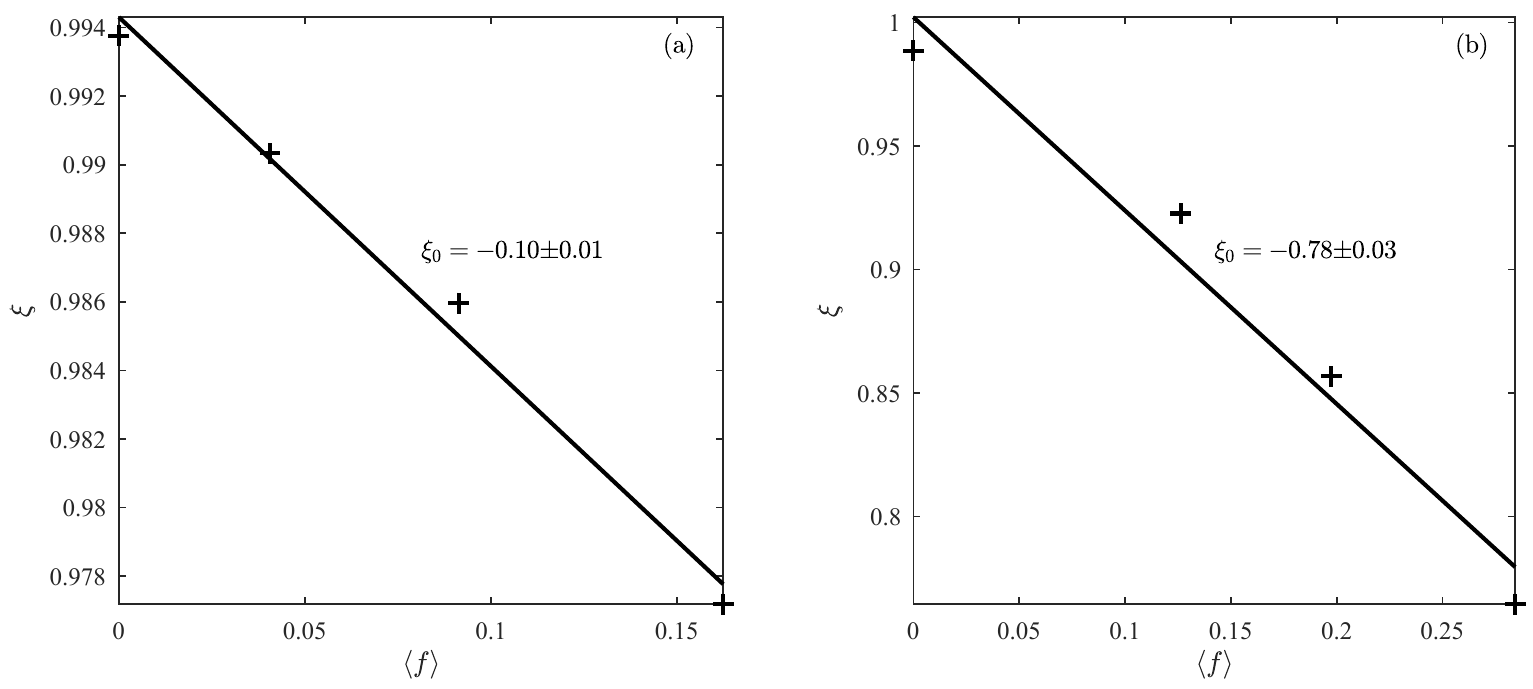}
\end{center}
\caption{\label{fig:a0study} Generalized relation between average curvature measured by the Ricci scalar and discrete energy levels for wrinkled (a) and bi-axialy rippled (b) graphene sheets. The crosses represent the data and the lines a linear fit with $\xi_0$ being the slope.}   
\end{figure*}

The relation in Eq.~(\ref{eq:llflat}), is verified for the flat sheet in the inset of Fig.~\ref{fig:FFT} where $E_n$ is plotted against $\sqrt{2Bn}$. Equivalently to the flat sheet the wrinkled graphene sheet follows the same functional dependence $E_n \approx sgn(n)\sqrt{2B|n|}$. The following relation between the energy eigenvalues and magnetic field is proposed,\cite{jd_paper}
\begin{equation}
\label{eq:llcurved}
    E_n=\xi(\langle f \rangle)\sqrt{2B|n|},
\end{equation}
where $\xi(\langle f \rangle)$ is a deformation dependent parameter where the average deformation is calculated from
\begin{equation}
       \langle f \rangle=\frac{1}{l_x}\int_0^{l_x} f(x)dx.
\end{equation}
The result is shown in Fig.~\ref{fig:a0study}(a). An identical study is also carried for the bi-axially rippled graphene sheets parameterized as out of plane deformations by
\begin{equation}
\label{eq:2Dripples}
    h(x,y)=a_0 \cos\bigg(\frac{2\pi k_0x}{l_x}\bigg)\cos\bigg(\frac{2\pi k_0y}{l_y}\bigg),
\end{equation}
where $l_y=l_x$, see Fig.~\ref{fig:geometries}
and Fig.~\ref{fig:a0study}(b). In conclusion the Landau levels in curved graphene sheets are given by
\begin{equation}
    E_n=(1+\xi_0(\langle f \rangle)\sqrt{2B|n|},
\end{equation}
where $\xi_0=-0.10 \pm 0.004$ and $\xi_0=-0.78 \pm 0.03$  
for wrinkled and bi-axialy rippled graphene respectively. Comparing with the preceding work, Ref.~\cite{jd_paper}, in the case of uni-axially rippled graphene sheets $\xi_0=-0.57 \pm 0.04$.

The effect can be understood as an effective magnetic field experienced from the curved graphene sheets. The total magnetic flux threading the sheet is
\begin{equation*}
\label{eq:flux}
\begin{split}
    \Phi & =\int_S \Vec{B}\cdot d \Vec{S}  =\iint \Vec{B}\cdot \Vec{n} \sqrt{g} dxdy     \\
  & = B l_x l_y= B \mathcal{A}_0,
\end{split}
\end{equation*}
where $l_x,l_y$ are the dimensions and $\mathcal{A}_0$ the area of the flat sheet. The total magnetic flux is independent of curvature as only the perpendicular projection between the magnetic field and surface contributes to the flux. By defining $\Phi=B \mathcal{A}_0=B_{eff} \mathcal{A}$, with $B_{eff}$ the effective field and $\mathcal{A}=\iint \sqrt{g}dxdy$, the true area of the curved sheet, $B_{eff}$ can be computed from 
\begin{equation*}
\label{eq:flux}
\begin{split}
    B_{eff} & = B \mathcal{A}_0 \mathcal{A}^{-1}= B \mathcal{A}_0 \bigg[\iint \sqrt{1+h_x^2}~dxdy \bigg]^{-1}\\
  & = B \mathcal{A}_0 \big[\mathcal{A}_0+\mathcal{A}_0 \langle f \rangle + \mathcal{O}(h_x^4 \sim a_0^4)\big]^{-1} \\
  & = B (1-\langle f \rangle)+\mathcal{O}(a_0^4).
\end{split}
\end{equation*}
When this is plugged into the energy level law, Eq.~(\ref{eq:llflat}),
\begin{align}
    E_n=sgn(n)\sqrt{2B_{eff}|n|} & \approx \sqrt{2(1-\langle f \rangle)B|n|} \\
    & \approx (1-0.5\langle f \rangle)\sqrt{2B|n|},
\end{align}
for small deformations. Comparing all three $\xi_0\approx 0.10, 0.57, 0.79$, for wrinkled, uni-axial and bi-axial rippled respectively, it is evident that, in the linear approximation, the Landau level shifting is around $\xi_0 \sim 0.5$  depending on the complexity of the curvature.

For larger deformations it can be expected that the higher order terms  become important enough such that a linear relation would no longer be valid. The current numerical method is limited, due to stability bounds, to deformations less than $15 \%$. The simulated result lies within the range of experimental validity. For a magnetic field $B \sim 1$T, the energy gap for flat graphene between $n=0$ and $n=1$, $E_1\sim 35$meV, as seen from Ref.~\cite{Sadowski2006}, while the shifting of the Landau levels would be about $13\%$ for $10\%$ deformation (i.e height of $\sim 20$nm). This is a significant enough effect to be observable.

\subsection{Resistivity of wrinkled graphene sheet}

\begin{figure*}
\includegraphics[width=2\columnwidth]{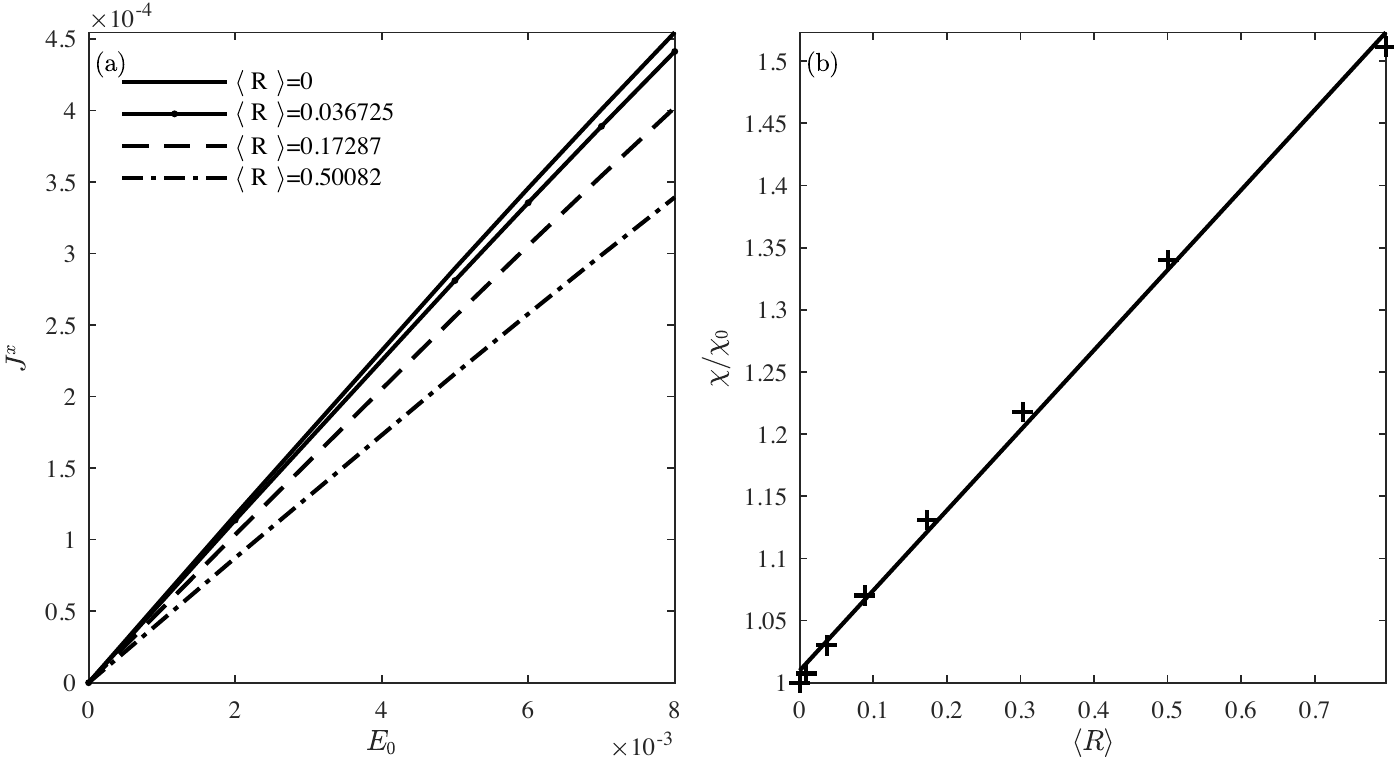}
\caption{\label{fig:resistivity} (a) Current, electric field relation for different average curvatures of wrinkled graphene sheets. The slope of the lines is the conductivity. (b) Resistivity against average curvature for wrinkled graphene sheets. The crosses represent the data and the solid line a linear fit.}
\end{figure*}

To investigate the conductivity relation for curved graphene sheets a square and periodic domain is initialized to $L_x \times L_y=128 \times 128$ with the same initial wavefunction as in Sec.~\ref{sec:LL}. Applying an electric field gradient along the $x$ direction, $E_x=E_0(x-x_0)$, a parallel current, $ J^x$, develops. The average current after 100 time steps is plotted against the electric field, as seen in Fig.~\ref{fig:resistivity}(a).  The current is linearly related to the  electric field, where the slope of the lines in Fig.~\ref{fig:resistivity}(a) is the conductivity. The effect of curvature, i.e. wrinkles in graphene, is to decrease the conductivity of the graphene sheet, smaller slope, in Fig.~\ref{fig:resistivity}(a). The resistivity $\chi$ can be calculated as the inverse of the conductivity from the relation $J^x=E_0/\chi$. The resistivity is linearly increasing with curvature according to 
\begin{equation}
    \frac{\chi}{\chi_0}=\zeta \langle R \rangle+c,
\end{equation}
where $\zeta =0.64 \pm 0.02$ and $c= 1.01 \pm 0.02$, as seen in Fig.~\ref{fig:resistivity}(b). The space averaged Ricci scalar $\langle R \rangle$:
\begin{equation}
\langle R \rangle= \bigg| \bigg( \int\limits^{x} R(x) \sqrt{g}dx\bigg) / \int\limits^{x} \sqrt{g}dx \bigg|, 
\end{equation}
is used as the measure of curvature. 

This result is qualitatively in agreement  to experimental investigations, as in Ref.~\cite{graphene_wrinkles_ripples}, where the resistivity of the graphene sheet decreases linearly when stretched into a flat sheet. 

\section{Conclusions and outlook}

Implementing the curved space QLB method we have shown that graphene Landau levels are shifted in the presence of curvature in strained graphene sheets. We have studied the cases of wrinkled and bi-axially rippled graphene.  The energy levels follow a similar square root dependence on the energy quantum number as for the rippled and flat graphene sheets. The Landau levels in wrinkled and rippled graphene sheets are given by $E_n=(1+\xi_0(\langle f \rangle)\sqrt{2B|n|}$, where $\xi_0$ is a geometry and curvature dependent parameter. The effect is higher for the more complex geometries and it can be verified experimentally with magnetic fields around $1$T. Additionally, the resistivity of wrinkled graphene sheets is shown to obey a linear relation with average space curvature, $\chi=\zeta \langle R \rangle$. 

The curved space QLB method opens up the possibility of exploring relativistic quantum particles and condensed matter systems on curved surfaces. This solver can be further developed to include time curvature and interaction terms, opening up the possibility of numerically solving complex and strongly coupled quantum field theories for solid state physics.

\section*{Acknowledgments}
The authors are grateful for the financial support Swiss National Science Foundation, under Grant No. 200021 165497 and HJH thanks CAPES for support.

\bibliographystyle{ieeetr}


\appendix
\end{document}